\documentclass{copernicus}

\usepackage{epstopdf, soul, color, ulem}
\begin{document}

\title{Incorporating magnetic field observations in wind models of low-mass stars}
\author[1,2]{A.~A.~Vidotto}

\affil[1]{School of Physics \& Astronomy, Univ.~of St Andrews, North Haugh, St Andrews, KY16 9SS, UK}
\affil[2]{Observatoire de Gen\`eve, Univ.~de Gen\`eve, Ch.~des Maillettes 51, Versoix, CH-1290, Switzerland}


\runningtitle{Incorporating magnetic field observations in models of stellar winds}
\runningauthor{A.~A.~Vidotto}
\correspondence{A.~A.~Vidotto (Aline.Vidotto@unige.ch)}

\received{}
\pubdiscuss{} 
\revised{}
\accepted{}
\published{}

\firstpage{1}

\maketitle  

\begin{abstract}
Stellar winds of cool, main-sequence stars are very tenuous and difficult to observe. Despite carrying away only a small amount of the stellar mass, they are important for regulating the rotation of the star and, consequently, its activity and magnetism. As it permeates the interplanetary space, the stellar wind interacts with any exoplanet encountered on its way, until it reaches the interstellar medium  (ISM). These interactions can result in complex physical processes that depend on the characteristics of the wind. To better constrain the wind characteristics, more realistic wind models that  account for factors such as stellar rotation and the complex/diverse observationally-derived stellar magnetic field configurations of cool stars are required. In this paper, I present a three-dimensional model of the wind of cool stars, which adopt as boundary condition observationally-derived magnetic maps. I also discuss how these studies are relevant for, e.g., the characterisation of the interaction between stellar winds and planets/ISM, and the propagation of cosmic rays.
\end{abstract}

\introduction  
\vspace{-0.2cm}
Magnetic fields play an important role along the stellar life. For low-mass stars, they are believed to regulate stellar rotation from the early stages of star formation until the ultimate stages of the life of a star. Empirically, the projected rotational velocities  $v \sin(i)$ of G-type stars in the main-sequence phase decrease with age $t$ as $v\sin(i) \propto t^{-1/2}$ \citep{1972ApJ...171..565S}. This rotational braking is believed to be caused by stellar winds, which, outflowing along magnetic field lines, are able to efficiently remove the angular momentum of the star \citep[e.g.,][]{1958ApJ...128..664P,1967ApJ...148..217W}. 

As the wind flows out of the star, it impinges on any exoplanet encountered on its way, all the way until it reaches the ISM. The interactions between the stellar wind and exoplanets/ISM can result in complex physical processes that depend on the  characteristics of the wind. These characteristics depend, among others, on the particular geometry of the stellar magnetic field. Modern techniques, such as Zeeman-Doppler Imaging (ZDI), have made it possible to reconstruct the large-scale surface magnetic fields of other stars \citep{2009ARA&A..47..333D}. This method has now been used to investigate the magnetic topology (i.e., intensity and orientation) of more than 100 low-mass stars. Thanks to this intense observational effort, a more general understanding of how the large-scale field correlates with age, activity and rotation is starting to emerge \citep{vidotto14}. In addition, given all this recent insight gained into the magnetic topology of different stars, we are now able to produce more realistic models of  winds of low-mass stars, directly incorporating the observationally derived stellar magnetism into three-dimensional (3D) numerical simulations.

\section{Numerical models of stellar winds}
\vspace{-0.2cm}
To account for the observed 3D nature of stellar magnetic fields, 3D stellar wind models are required. Here I present the technique used in \citet{2011MNRAS.412..351V,2012MNRAS.423.3285V,2014MNRAS.438.1162V,2013MNRAS.431..528J, 2013MNRAS.436.2179L}, which incorporates the observationally reconstructed magnetic fields using the ZDI technique. The wind models were simulated using the 3D magnetohydrodynamics (MHD) numerical code BATS-R-US \citep{1999JCoPh.154..284P}. BATS-R-US solves the ideal MHD equations
\begin{equation}
\label{eq:continuity_conserve}
\frac{\partial \rho}{\partial t} + \boldsymbol\nabla\cdot \left(\rho {\bf u}\right) = 0,
\end{equation}
\begin{equation}
\label{eq:momentum_conserve}
\frac{\partial \left(\rho {\bf u}\right)}{\partial t} + \boldsymbol\nabla\cdot\left[ \rho{\bf u\,u}+ \left(P + \frac{B^2}{8\pi}\right)I - \frac{{\bf B\,B}}{4\pi}\right] = \rho {\bf g},
\end{equation}
\begin{equation}
\label{eq:bfield_conserve}
\frac{\partial {\bf B}}{\partial t} + \boldsymbol\nabla\cdot\left({\bf u\,B} - {\bf B\,u}\right) = 0,
\end{equation}
\begin{eqnarray}
\label{eq:energy_conserve}
&& \frac{\partial}{\partial t} \left[\frac{\rho u^2}{2}+\frac{P}{\gamma-1}+\frac{B^2}{8\pi}\right]  + \nonumber \\
& \boldsymbol\nabla& \cdot \left\{ {\bf u} \left[ \left( \frac{\rho u^2}{2}+\frac{P}{\gamma-1}+\frac{B^2}{8\pi} \right) + P + \frac{B^2}{8\pi} \right] - \frac{\left({\bf u}\cdot{\bf B}\right) {\bf B}}{4\pi}\right\}  \nonumber\\
&=& \rho {\bf g}\cdot {\bf u} ,
\end{eqnarray}
 at the inertial reference frame, where the eight primary variables are the mass density $\rho$, the plasma velocity ${\bf u}$, the magnetic field ${\bf B}$, and the gas pressure $P$. The gravitational acceleration is given by ${\bf g}$, and $\gamma$ is the polytropic index ($p\propto \rho^\gamma$). We consider an ideal gas, so $P=n k_B T$, where  $k_B$ is the Boltzmann constant, $T$ is the temperature, $n=\rho/(\mu m_p)$ is the particle number density of the stellar wind, $\mu m_p$ is the mean mass of the particle. 

\uline{Simulation grid:} Our grid is Cartesian and, for the simulations we illustrate next, the grid extends in $x$, $y$, and $z$ from $-20$ to $20$~stellar radii ($R_\star$). The star is placed at the centre of the grid. The finest resolved cells (resolution of $0.0097~R_\star$) are located close to the star and the coarsest cells  ($0.31~R_\star$) are located at the outer edges of the grid. The total number of cells in our simulations is $\sim80$ million.

{\uline{Boundary conditions:} The inner boundary of our simulation is located at $r=R_\star$, where the radial component of the magnetic field ($B_r$) is set to be the observationally-reconstructed one. $B_r$ is held fixed at the boundary throughout the simulation run, as are $\rho$ and $P$. A zero radial gradient is set to the remaining components of ${\bf B}$ and ${\bf u}=0$ in the frame co-rotating with the star. At the outer boundaries (at the edges of the grid), a zero gradient is set to all the primary variables. The rotation axis of the star is aligned with the $z$-axis, and the star is assumed to rotate as a solid body.

\uline{Initial state:} At the initial state of the simulations, the wind profile is that of a non-magnetised, thermally-driven wind \citep{1958ApJ...128..664P}. The magnetic field considered in the grid is derived from extrapolations of observed surface radial magnetic maps using the potential-field source surface method \citep{1969SoPh....9..131A}. Starting from an initial potential magnetic field configuration and a thermally-driven wind, the system is then evolved in time, resulting in a self-consistent interaction of the wind particles and the magnetic field lines. 

{The resultant solution, obtained self-consistently, is found when the system reaches steady state in the reference frame co-rotating with the star.} The left (right) panel of Figure~\ref{fig1} illustrates the initial (final) configuration of the magnetic field lines in the wind simulation of DT~Vir, based on the simulations presented in \citet{2014MNRAS.438.1162V}.  {Note that, because of the interaction between particles and magnetic field lines, currents are created in the system and the magnetic field is removed from its initial potential configuration, becoming stressed.} DT~Vir is an early-M dwarf star, with a mass of $0.59~M_\odot$, radius $0.53~R_\odot$ and rotation period of $2.85~$d. {We assume $\gamma=1.1$ and $\mu=0.5$. The other free parameters of our model are the wind base temperature ($2\times10^6$~K) and the density ($10^{11}$~g~cm$^{-3}$). Ideally, we would like to constrain these parameters from observations of  winds of cool stars, but these are rather challenging to obtain and no wind constraints exist for DT~Vir. On the other hand, X-ray observations can provide estimates on the coronal temperatures and densities} \citep{2004A&ARv..12...71G}, {helping to constrain the densities and temperatures at the wind base. The values adopted for DT~Vir are within the observed range provided by X-ray observations.}

\begin{figure}
\begin{center}
\includegraphics[width=5.5cm]{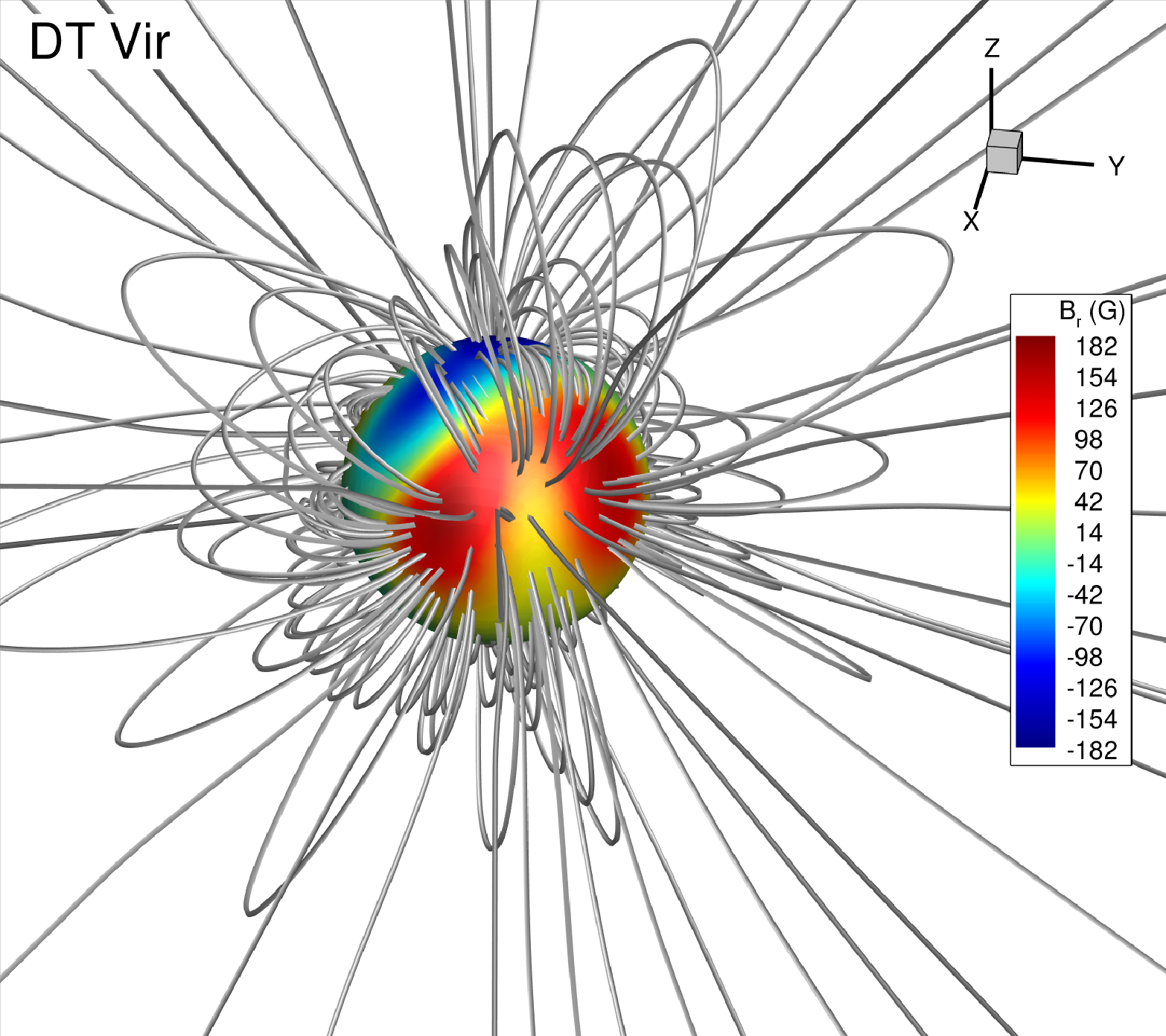}
\includegraphics[width=5.5cm]{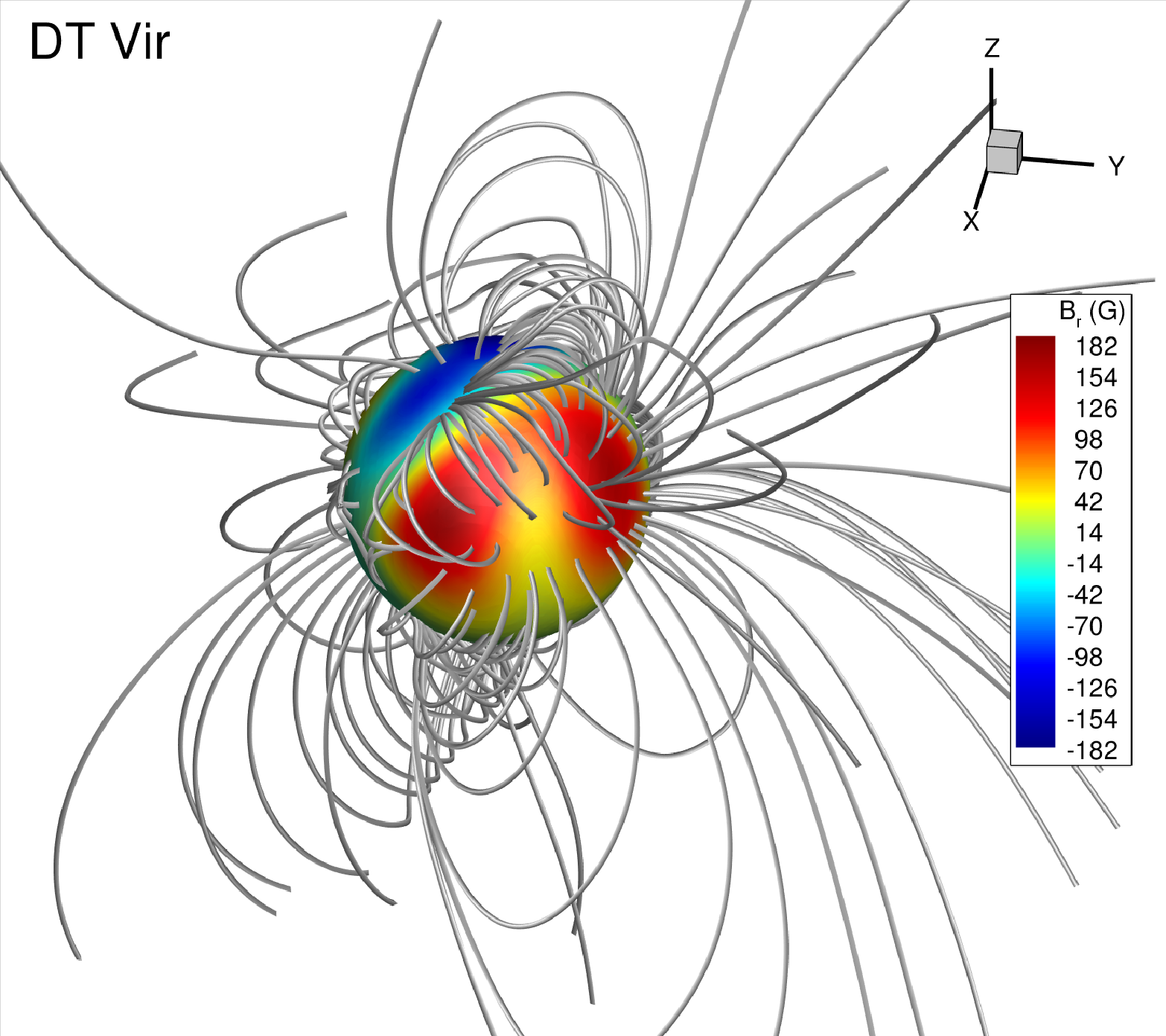}
\end{center}
\caption{Left: The initial configuration of the magnetic field lines extrapolated from observed surface radial magnetic maps (colour-coded, from \citealt{2008MNRAS.390..545D}) assuming a potential field. Right: Final configuration of the magnetic field, derived from self-consistent interaction between wind particles and magnetic field lines.} \label{fig1}
\end{figure}

In total, \citet{2014MNRAS.438.1162V} performed numerical simulations of the winds of a sample of $6$ early-M dwarfs whose magnetic fields were observationally reconstructed \citep{2008MNRAS.390..545D}.They showed that the wind mass flux is essentially modulated by the local value of $|B_r|$, which share roughly the same modulation as the surface $|B_r|$ (observed). This implies that the more non-axisymmetric topology of the stellar magnetic field results in more asymmetric mass fluxes.

\section{A few comments on the relevance of stellar wind studies}
\vspace{-0.2cm}
The study of the local characteristics of the stellar wind is important for the characterisation of complex interactions that take place between exoplanets and their host stars. Planetary magnetospheric sizes are roughly set by pressure equilibrium between the planet's magnetic field and {the stellar wind total pressure $p_{\rm tot}$ (i.e., the sum of thermal, magnetic and ram pressures).} \citet{2014MNRAS.438.1162V} found that, close to the star,  $p_{\rm tot}$  is modulated by $|B_r|$ and the more non-axisymmetric topology of the stellar magnetic field produces more asymmetric distributions of $p_{\rm tot}$. As an exoplanet orbits around its host star, it probes regions of different $p_{\rm tot}$. Consequently, its magnetospheric size becomes smaller (larger) when the external $p_{\rm tot}$ is larger (smaller). In addition to magnetospheric size variations, the orientation of the bow shock that forms surrounding planetary magnetospheres also changes along the planetary orbit, as a consequence of the asymmetric nature of stellar magnetic fields.  Other effects, such as planetary obliquity, eccentricity, and  temporal variations of the stellar magnetic field, can also cause variability in planetary magnetospheric characteristics \citep{2011MNRAS.414.1573V}.

The stellar wind can also affect the distances and shapes of ``astropauses'', defined as the surface where the pressure of the stellar wind balances the pressure of the ISM.  In the presence of highly asymmetric stellar winds, which could be caused by the non-axisymmetric nature of the stellar magnetic field, astropauses might also lack symmetry. This can be a long-lasting phenomena or occur only at certain epochs of a magnetic cycle. {Indeed, models of the interaction zone between the solar wind and the ISM suggests that the heliopause can be influenced due to solar cycle effects} \citep{2014ApJ...783L..16B}. {In the case of the solar system, the shape of the termination shock is likely more strongly influenced than the shape of the heliopause } \citep[e.g.,][]{2003GeoRL..30.1045S} {and this might also be the case in stellar systems.}

Finally, the geometry of B and the global characteristics of stellar winds can also affect propagation of cosmic rays (CRs). In the case of the Earth, the galactic CR flux is modulated over the solar cycle, being inversely correlated with the non-axisymmetric component of the total open solar magnetic flux \citep{2006ApJ...644..638W}. By analogy, if the non-axisymmetric component of the stellar magnetic field is able to reduce the flux of CRs reaching an exoplanet, then we would expect that planets orbiting stars with largely non-axisymmetric fields would be more shielded from galactic CRs, independently of the planet's own shielding mechanism (such as the ones provided by a thick atmosphere or large magnetosphere). {The star can also be an important source of CRs. In the case of active stars, intensive fluxes of energetic electrons might remain trapped in the closed magnetic field structures. Close-in exoplanets that cross these large-scale structures might be exposed to intensive stellar CRs. In this case, the analogy regarding the non-axisymmetric component of B, which anti-correlates with the CRs flux on Earth, might not be valid. More study on that issue is needed.}

\vspace{-0.6cm} 
\begin{acknowledgements}
I would like to thank the organisers for inviting me to attend this workshop and for contributing towards my travel expenses. I also thank the referees (Dr T.~Wiengarten and an anonymous one) who provided constructive feedbacks during the preparation of this manuscript. I acknowledge support from a Royal Astronomical Society Fellowship and an Ambizione Fellowship from the Swiss National Science Foundation.
\end{acknowledgements}

\def\aj{{AJ}}                   
\def\araa{{ARA\&A}}             
\def\apj{{ApJ}}                 
\def\apjl{{ApJ}}                
\def\apjs{{ApJS}}               
\def\ao{{Appl.~Opt.}}           
\def\apss{{Ap\&SS}}             
\def\aap{{A\&A}}                
\def\aapr{{A\&A~Rev.}}          
\def\aaps{{A\&AS}}              
\def\azh{{AZh}}                 
\def\baas{{BAAS}}               
\def\jrasc{{JRASC}}             
\def\memras{{MmRAS}}            
\def\mnras{{MNRAS}}             
\def\pra{{Phys.~Rev.~A}}        
\def\prb{{Phys.~Rev.~B}}        
\def\prc{{Phys.~Rev.~C}}        
\def\prd{{Phys.~Rev.~D}}        
\def\pre{{Phys.~Rev.~E}}        
\def\prl{{Phys.~Rev.~Lett.}}    
\def\pasp{{PASP}}               
\def\pasj{{PASJ}}               
\def\qjras{{QJRAS}}             
\def\skytel{{S\&T}}             
\def\solphys{{Sol.~Phys.}}      
\def\sovast{{Soviet~Ast.}}      
\def\ssr{{Space~Sci.~Rev.}}     
\def\zap{{ZAp}}                 
\def\nat{{Nature}}              
\def\iaucirc{{IAU~Circ.}}       
\def\aplett{{Astrophys.~Lett.}} 
\def\apspr{{Astrophys.~Space~Phys.~Res.}}   
\def\bain{{Bull.~Astron.~Inst.~Netherlands}}    
\def\fcp{{Fund.~Cosmic~Phys.}}  
\def\gca{{Geochim.~Cosmochim.~Acta}}        
\def\grl{{Geophys.~Res.~Lett.}} 
\def\jcp{{J.~Chem.~Phys.}}      
\def\jgr{{J.~Geophys.~Res.}}    
\def\jqsrt{{J.~Quant.~Spec.~Radiat.~Transf.}}   
\def\memsai{{Mem.~Soc.~Astron.~Italiana}}   
\def\nphysa{{Nucl.~Phys.~A}}    
\def\physrep{{Phys.~Rep.}}      
\def\physscr{{Phys.~Scr}}       
\def\planss{{Planet.~Space~Sci.}}           
\def\procspie{{Proc.~SPIE}}     
\def\actaa{{Acta~Astronomica}}     
\def\icarus{{ICARUS}}

\let\astap=\aap
\let\apjlett=\apjl
\let\apjsupp=\apjs
\let\applopt=\ao
\let\mnrasl=\mnras


\end{document}